\documentstyle[twocolumn,aps,floats,epsf]{revtex}

\begin{document}
\twocolumn[
\hsize\textwidth\columnwidth\hsize\csname@twocolumnfalse\endcsname
\draft
\title{Unexpected behavior in $^3$He films and the
theory of fermions in two dimensions}
\author{Paulo F. Farinas}
\address{Instituto de F\' \i sica Gleb Wataghin,
Universidade Estadual de Campinas 13083-970,
Campinas, S\~ ao Paulo, Brazil}

\date{\today}
\maketitle

\begin{abstract}
It is shown that the recently observed and unexplained behavior
for the spin diffusion in $^3$He films can be
understood in terms of the Fermi liquid theory,
provided nonperturbative effects characteristic
of the two dimensions are taken into account.
It is conjectured that such a behavior suggests the
onset of a regime in which spin-current is conserved.

\end{abstract}
]

The theory of Fermi liquids in two dimensions ($2d$) has been
at the center of various settlements involving unexpected
physics that has been arising in a broad range of interesting
materials. Examples range from systems in the quantum Hall regime
to copper-oxide superconductors, including quantum-fluid mixtures
of $^3$He-$^4$He. These latter provided grounds for
a recent observation
of unexpected behavior, whose main cause will be argued in the
following to be a consequence of the two dimensionality
of a general system of fermions.
In a recent Letter,\cite{hall1} Sheldon and Hallock presented
experimental results for spin diffusion in a thin film of $^3$He
on top of a superfluid $^4$He film. The
result for the spin-diffusion coefficient ($D$) as
a function of the $^3$He coverage ($D_3$), shows an upturn
followed by a saturation and a downturn, as $D_3$ is increased.
These features could not be reproduced by standard theoretical
predictions.\cite{mula}

Throughout their past
work, Hallock and collaborators have
managed to develop a notable degree of controll over
the various regimes that may be set in the
rich systems that constitute $^3$He-$^4$He films.\cite{Hallock}
In the work reported in Ref.\cite{hall1}, care has been taken
to guarantee that the $^3$He film forms a $2d$ system of
interacting fermions, with little
influence from the $^4$He superfluid film
that seats below. Hence one should
be able to account for the observed features
using theoretical results from a $2d$
Landau theory. 

It is argued in this article that the theory
of $2d$ Fermi liquids
{\it does} account for the
observed behavior (regarding the existence
of a peak in $D$ as a function of $D_3$)
provided one includes
nonperturbative effects that are
inherent to a Fermi liquid
in two spatial dimensions.\cite{jan1,remk1}
It is also pointed
that the onset of
regimes in the $2d$ Fermi liquid
for which spin-current is a conserved quantity,\cite{us}
is related to such a peaked behavior, however
more experimental results are necessary to establish
whether these regimes may be tuned or not.

\begin{figure}[h]
\epsfxsize=4.3in
\leftline{\epsfbox{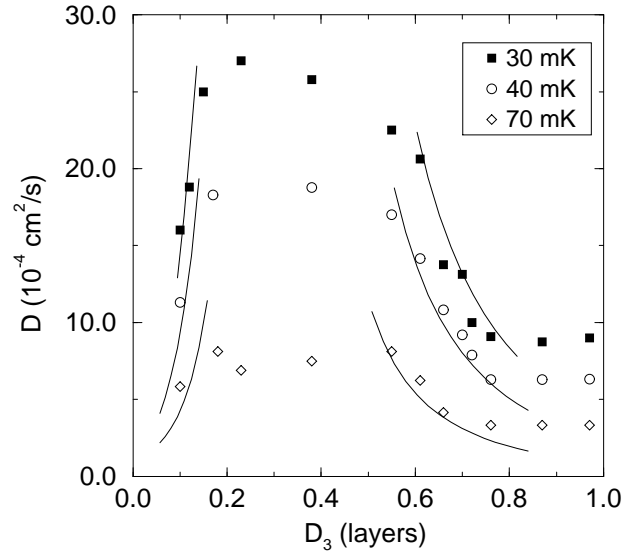}}
\vspace{-7.5truecm}
\renewcommand{\baselinestretch}{1.1}
\caption{Experimental data of Sheldon and Hallock
(symbols) and the theoretical fits (solid lines)
at the corresponding temperatures; same
values of two fitting parameters
are used for all temperatures: the effective hard-core
radius $a_s = 0.75$ {\AA } whilst the vertical amplitude
is fixed using solely the first experimental
point at 30mK.
\label{fig1}}
\end{figure}

I shall start with some remarks on existing theoretical results, and
explain why the first attempts to fit the experimental data failed.
I shall then devise the microscopic scenario that accounts
for the physics underlying the peaked behavior observed,
and calculate $D$, whose result
is plotted in Fig.(\ref{fig1}), compared to the experimental points.
It should be stressed that the aim here is not to fully
reproduce the experimental data, but yet uncover the physics
that leads to the observed peak. 
Indeed, although the accuracy of the theoretical
fitting is quite remarkable
and consistent with the long withstanding experimental
observations of Owers-Bradley {\it et al.}\cite{ob}
the existence of a saturation plateau is not accountable
in the simple model studied here. One possible reason for such
a behavior is discussed towards the end of the text.

The phenomenological result
for the spin-diffusion coefficient in two dimensions
was derived in Ref.\cite{mula} and reads

\begin{equation}
D={\hbar \over {\pi m^*}}{{(1+F_0^a)^3C(\lambda )}%
 \over {
(F_0^a)^2(T_F/T)^2ln(T_F/T)}} 
\; , \label{diff}
\end{equation}
where $F_\ell^{a(s)}$ stands for the usual Landau
antisymmetric (symmetric) parameters, $m^*$ is the
effective mass, $T_{(F)}$ is the (Fermi) temperature,
and $C(\lambda )$ is a function of all Landau
parameters, given by Eq.(\ref{c}). On fitting
Eq.(\ref{diff}) to experimental data, one
should be careful with the approximation
that is taken for $C(\lambda )$. Based on the {\it strictly
pertubative} results of Havens-Sacco and Widom,\cite{saco}
one may reasonably disregard all high order Landau parameters,
which leads to an approximately constant value for
$C(\lambda )$. Hallock and Sheldon tried to fit their
experimental results using this particular approach, with
$C(\lambda )\sim 3/4$.
However, the existence of nonperturbative effects intrinsic
to the $2d$ Fermi liquid, predicted some time ago by Engelbrecht
and Randeria,\cite{jan1} suggests that such an approximation
for the diffusion coefficient is not valid in general. Although
single Landau parameters, which are present in the formulas
of most the equilibrium properties, may not show any
unusual behavior as a function of the density,
the combination of various (usually infinite)
Landau parameters may drive a $2d$ Fermi liquid into
quite exotic regimes when the density is varied,
as it has been pointed recently.\cite{us,us2}

To carry a more specific discussion, the full expression
for $C(\lambda )$,

\begin{equation}
C(\lambda )=\frac{\lambda (A_0^a)^2}%
{4N^2(0)z^4}\sum_{n=1}^{\infty}%
{{(4n-1)/(2n^2-n)}\over {2n^2-n-%
1 + \lambda |\Gamma_{\uparrow %
\downarrow \downarrow \uparrow}^{k}(\pi )|^2%
}}\; , \label{c}
\end{equation}
is conveniently written in terms of the
vertex function
of perturbation theory taken in the limit of zero
frequency,\cite{dial}
$\Gamma_{\sigma_1\sigma_2\sigma_3\sigma_4}%
^{k}(\theta )$,
where $\theta $ is the angle between the momenta
of two
quasiparticles in a scattering process constrained
by a Fermi circle and $\sigma_i$ are spin labels.
$z$ and $N(0)$ are respectively the Green's function residue 
and the density of states, both at the Fermi surface.
Here $\lambda $ stands for the inverse of a phase space
sum,\cite{remk2}
\[
\lambda = \frac{8}{\sum_{\sigma_i;\theta = 0,\pi }%
|\Gamma_{\sigma_1\sigma_2\sigma_3\sigma_4}%
^{k}(\theta )|^2\delta_{\sigma_1+\sigma_2,%
\sigma_3+\sigma_4}}.
\]
It is a straightforward yet not pleasant task 
to obtain these expressions from the results of Ref.\cite{mula}
This convenience of having closed expressions in
terms of microscopic quantities is a remarkable
amenability of the theory in two dimensions.
The projection of $\Gamma_{\sigma_1\sigma_2\sigma_3\sigma_4}%
^{k}(\theta )$ on a conveniently chosen basis
defines the Landau scattering amplitudes $A_\ell^{a(s)}$.\cite{baym}
Using the forward scattering sum rule
for these amplitudes, $\sum_\ell A_\ell^{a}+A_\ell^{s}=0$,
one readily sees that if only the leading
order $A_0^{a(s)}$ are kept then $C(\lambda )$ is
constant.\cite{remk4}
Note, however, that varying the density may
in principle drive the Landau coefficients to combine
in such a way as to bring $\Gamma_{\uparrow %
\downarrow \downarrow \uparrow}^{k}(\pi )$
down to small
values, which causes $D$ to increase.
In this particular case,
if the spin-diffusion relaxation time becomes very
large, either another diffusion mechanism as spin-lattice
relaxation sets in or, in the ideal $2d$ interacting
liquid, the system will enter a regime for which spin-current
is conserved and a spin-viscous damping mechanism becomes
associated with the transport of magnetization throughout the
sample.\cite{us,us2} A sharp raising of the spin-diffusion
coefficient when one varies the density
is then one of the signatures expected at the onset of such
a regime.

In order to obtain a result that includes all the
parameters, I start with the expression obtained by
Engelbrecht, Randeria, and Zhang\cite{jan2} 
for the vertex part in the limit
of zero exchange momentum,

\begin{eqnarray}
\Gamma_{\uparrow %
\downarrow \downarrow \uparrow}%
^{\omega }(\theta )=\frac{\pi \hbar^2}%
{z^2m_h}\biggl\{ 2\pi n a_s^2&&\sin^2(\theta /2)
\biggr.
\nonumber \\
\biggl. &&+\frac{1}{\ln \left[2\pi n a_s^2%
\cos^2(\theta /2)\right]}\biggr\}
\; , \label{gamma}
\end{eqnarray}
where $m_h$ is the $^3$He hydrodynamic mass, $a_s$ is
a hard-core potential radius, and $n$ is the density,
which is linearly related to the layer coverage $D_3$
in the experiment of Ref.\cite{hall1}

The expansion paramenter of the theory is $g=-(\ln 2\pi %
n a_s^2)^{-1}$, and for coverages ranging from $0 - 0.8$,
$g$ falls within the interval $0 - 0.58$. Equation (\ref{gamma})
follows from an all-order ladder sumation in the
spirit of the long known results in three
dimensions.\cite{Galitskii} In this sense, both terms
on the right side of Eq.(\ref{gamma}) are nonperturbative.
The first term arises from a pole in the vertex
function below the particle-hole
continuum, and is a particularity of two dimensions,
since it results from the finite value of the density
of states for small momenta. This pole is associated
with a two-hole bound state.\cite{jan1}
Note that this pole term has an essential sigularity
in $g$ and hence it cannot be determined to any finite
order in perturbation theory.
The second term contains the particle-hole continuum
contributions. Although these contributions
are separable in the dilute limit
defined by the small values of $g$, the existence of a pole
also phase-shifts the continuum and on determining the
unusual behavior in the spin-diffusion coefficient, {\it
both} terms are necessary.
However, the continuum alone yields no new physics in
two dimensions,
it is the counterpart of the known three dimensional
rendered logarithmic divergences
in the repulsive channel.\cite{remk3} Indeed, a calculation
using only the continuum (which will be shown elsewhere)
leads to the usual behavior for $D$, consistent with the
result of Ref.\cite{saco} in second order perturbation, and
also with Eq.(\ref{diff}) for a constant $C(\lambda )$.
This discussion establishes the two dimensional
nonperturbative nature of the unusual behavior 
in $D$, as derived in what follows.
\\

\noindent
In order to calculate the spin-diffusion coefficient
from Eq.(\ref{diff}) knowledge of the zero frequency
limit of the vertex part is necessary. These different
limits of the vertex function are related to each other
through a Bethe-Salpeter equation,\cite{dial}

\begin{eqnarray}
\Gamma^{k }(\theta )=\Gamma^{\omega }(\theta )
-N(0)z^2\int_0^{2\pi }\frac{d\theta' }{2\pi }
\Gamma^{\omega }(\theta - \theta')\Gamma^{k }(\theta' )
\; , \label{bethe}
\end{eqnarray}
where the spin indices have been supressed for convenience.
Using Eq.(\ref{gamma}) to define the kernel plus inhomogeneous
term, this equation is numerically solved and the result
used to calculate $D$ from Eq.(\ref{diff}). The outcome
is shown in Fig.(\ref{fig1}) To fit the experimental data
of Sheldon and Hallock at the given temperatures, a choice
of $a_s=0.75$ {\AA } was made. Although the width of the
peak is rather sensitive to this choice, it should be
noted that this value for $a_s$ is fairly close to the
long withstanding theoretical fits to the magnetic susceptibility
measuraments of Owers-Bradley {\it et al.}
in $^3$He films on Grafoil.\cite{ob} Such fittings
were achieved by using $a_s\sim 0.68$ {\AA } for a
hard-core potential and $a_s\sim 0.73$ {\AA } for a
6-12 potential.\cite{saco}
To account for the vertical scale, the theoretical
result was multiplied by a constant fixed by making the
first experimental point at $D_3=0.1$ for the $30$ mK
curve coincide with the theoretical value of
$D$ at the same coverage
and temperature.
The theoretical curves for all other
temperatures follow {\it with no extra adjustments}.
This fact, plus the consistency of the value for the
hard-core radius with the experimental result for the
magnetic susceptibility, which is an equilibrium response
in the spin channel, are rather suggestive that the physics
underlying the peaked behavior comes from the particularities
of interacting fermions in a reduced geometry.
\\
\begin{figure}[h]
\epsfxsize=4.3in
\leftline{\epsfbox{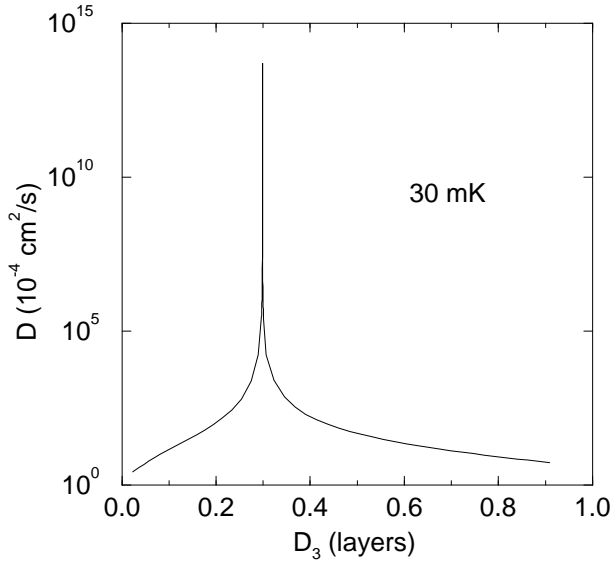}}
\vspace{-7.5truecm}
\renewcommand{\baselinestretch}{1.1}
\caption{Diffusion coefficient
of Fig.(\ref{fig1}) in a logarithmic scale.
\label{fig2}}
\end{figure}

The discontinuity in the derivative
of $D$ near
$D_3=0.8$, and its nearly constant behavior for intermediate
coverages are two features left unexplained by the
above considerations.
While the former is consistent with the promotion
of $^3$He atoms to the first excited surface state of the superfluid
$^4$He, as it is dicussed by Sheldom and Hallock in Ref.\cite{hall1}
the latter needs further considerations: A full range
plot of the theoretical result is shown in Fig.(\ref{fig2})
for $T=30$ mK. The sharp peak shown in this figure is
actually a divergence of $D$ occurring for $D_3\sim 0.3$
layers. This result holds for the {\it ideal} $2d$ Fermi liquid
and indicates that a regime for which spin-current is conserved
exists for the densities near the divergence.
However, the quasiparticles
are coupled to the Nuclepore substrate used in the
experiment. This complication is not treated
here, and it causes the spin-diffusion to be several
times smaller than the theoretical estimates (a mean field
tortuosity factor, independent of the coverage, seems to
be sufficient to correct the measured values).\cite{hall1}
Hence, the
spin-diffusion cannot grow indefinetly in such a non-ideal
system. The fact that for intermediate values of the
coverage, $D$
remains nearly constant as the coverage is varied
seems to be consistent with the nearly
constant ``spin-lattice'' relaxation times observed, suggesting
that a saturation-like behavior might be coming from
an underlying spin-relaxation mechanism with origin
not in the $2d$ system of fermions but rather in the
environment.

In summarizing, an explanation has been presented
for the existence of a peak in the recent measuraments
of the spin-diffusion coefficient as a function of the
density in $^3$He
films. Although the physical scenario builts from quite
traditional results, it is conjectured that
such a behavior might be signalizing the existence of
new regimes of the $2d$ Fermi liquids predicted recently
for which spin-current is conserved. The experimental
results are shown also to support the predictions of
Engelbrecht and Randeria for the existence of a collective
mode in systems of interacting fermions,
which exists only in two dimensions.

The author aknowledges useful
discussions with K.S. Bedell and
C. K\" ubert, the hospitality of A. Caldeira
at IFGW-UNICAMP where this work has been developed,
and N. Studart for quickly bringing the online
version of Ref.\cite{hall1} to his attention.
This work was supported in part by FAPESP, S\~ ao
Paulo, Brazil, Grant 98/07433-8.


\vspace{-0.5truecm}

\begin{references}
\vspace{-1.0truecm}

\bibitem{hall1}  P.A. Sheldon and R.B. Hallock, Phys. Rev.
Lett. {\bf 85}, 1468 (2000).

\bibitem{mula}K. Miyake and W.J. Mullin, Phys. Rev. Lett.
{\bf 50}, 197 (1983); J. Low Temp. Phys. {\bf 56}, 499 (1984).

\bibitem{Hallock} R.B. Hallock, Phys. Today, pg. 30
(June 1998) and references therein.

\bibitem{jan1}J. Engelbrecht and M. Randeria, Phys. Rev. Lett.
{\bf 65}, 1032 (1990).

\bibitem{remk1} Two dimensions means here that the quasiparticle's
momenta are confined to a surface (the fermi gas seats on a
surface state adsorbed on superfluid $^4$He); this leads
to a one dimensional Fermi surface.

\bibitem{us}  P.F. Farinas, K.S. Bedell, and N. Studart,
Phys. Rev. Lett. {\bf 82}, 3851 (1999).

\bibitem{ob}  J.R. Owers-Bradley, B.P. Cowan, M.G. Richards,
and A.L. Thomson,
Phys. Lett. {\bf 65A}, 424 (1978).

\bibitem{saco}S.M. Havens-Sacco and A. Widom, J. Low Temp.
Phys. {\bf 40}, 357 (1980).

\bibitem{us2}P.F. Farinas, K.S. Bedell, and N. Studart,
Physica B {\bf 280}, 93 (2000).

\bibitem{dial}A.A. Abrikosov, L.P. Gor'kov, and I.E. Dzyaloshinski,
{\it Methods of Quantum Field Theory in Statistical Mechanics}
(Dover, New York, 1975).

\bibitem{remk2} Note that the
quantity expressed here by $\lambda $
differs from that of Ref.\cite{mula} ($\lambda_-$).
They are related by $\lambda_- =
1 - \lambda |\Gamma_{\uparrow%
\downarrow \downarrow \uparrow}^{k}(\pi)|^2$.

\bibitem{baym}  G. Baym and C. Pethick, {\it Landau Fermi-Liquid
Theory} (Wiley, New York, 1991).

\bibitem{remk4} The precise value for this constant is easy to
calculate: $C_0(\lambda )=\pi^2/12$,
which is not much different from $3/4$.

\bibitem{jan2}J. Engelbrecht and M. Randeria, and
L. Zhang, Phys. Rev. B
{\bf 45}, 10135 (1992).

\bibitem{Galitskii}V.M. Galitskii, Zh. Eksp. Teor. Fiz. {\bf 34},
151 (1958); Sov. Phys. JETP {\bf 7}, 104 (1958).

\bibitem{remk3} These are the known singularities that, in
the attractive channel, signalize
the breakdown of the Fermi liquid scenario towards a BCS state.

\end{references}
\end{document}